\documentclass[manuscript]{emulateapj}

\usepackage{amsmath}

\newcommand{\mr}{\mathrm} %Roman type

\newcommand{\sgmg}{\Sigma_\mr{gas}} %surface gas density
\newcommand{\sgms}{\Sigma_\mr{star}} %surface stellar density
\newcommand{\sgmsf}{\Sigma_\mr{SFR}} %surface star formation rate density
\newcommand{\sgmin}{\Sigma_\mr{in}} %surface inflow rate density
\newcommand{\sgmout}{\Sigma_\mr{out}} %surface outflow rate density
\newcommand{\ret}{\mathcal{R}} %Return fraction
\newcommand{\tr}{(t,\:R)} %(t, R)

\newcommand{\kin}{k_\mr{in}} %kin
\newcommand{\etain}{\eta_\mr{in}} %etain
\newcommand{\kout}{k_\mr{out}} %kout
\newcommand{\etaout}{\eta_\mr{out}} %etaout

\newcommand{\ms}{M_\mr{cl}} %stellar mass of a star cluster
\newcommand{\mw}{m_\mr{w}} %outflow from a star cluster
\newcommand{\vw}{v_\mr{w}} %wind velocity

\shortauthors{TOYOUCHI \& CHIBA.}
\shorttitle{Gas Inflow and Outflow Histories in Disk Galaxies}

\begin{document}

\title{Gas Inflow and Outflow Histories in Disk Galaxies as Revealed from Observations of Distant Star-Forming Galaxies}

\author{Daisuke~Toyouchi\altaffilmark{1} and
	Masashi~Chiba\altaffilmark{1}}

\altaffiltext{1}{Astronomical Institute, Tohoku University,
Aoba-ku, Sendai 980-8578, Japan}

%%%%%% Abstract %%%%%%%%%%%%%%%%%%%%%%%%%%%%%%%%%%%%%%%%%%
\begin{abstract}
We investigate gas inflow and outflow histories in Milky Way-like disk galaxies, to get new insights into the baryonic processes in galaxy formation and evolution. For this purpose, we solve the equations for the evolutions of the surface mass densities of gas and metals at each radius in a galactic disk, based on the observed structural properties of distant star-forming galaxies, including the redshift evolution of their stellar mass distribution, their scaling relation between the mass of baryonic components, star formation rate (SFR) and chemical abundance, as well as the supposed evolution of their radial metallicity gradients (RMGs). We find that the efficiency of gas inflow for a given SFR decreases with time and that the inflow rate is always nearly proportional to the SFR. For gas outflow, although its efficiency for a given SFR is a decreasing function of time, similarly to gas inflow, the outflow rate is not necessarily proportional to the SFR and the relation between the outflow rate and SFR strongly depends on the evolution of the adopted RMG. We also find that the results on the outflow rate can be reproduced in the framework of momentum-driven (energy-driven) wind mechanism if the RMG is steepening (flattening) with time. Therefore if the well measured RMGs and their evolution for Milky Way-like galaxies are obtained from future observations, then our results will be useful to constrain the main driving mechanism for their galactic outflows.

\end{abstract}
%%%%%%%%%%%%%%%%%%%%%%%%%%%%%%%%%%%%%%%%%%%%%%%%%%%%%%%%%%


%%% Sec.1 %%%%%%%%%%%%%%%%%%%%%%%%%%%%%%%%%%%%%%%%%%%%%%%%
\section{INTRODUCTION}
Galactic inflow and outflow processes provide a significant influence on the budget of baryon and star formation activities in galaxies and thus play an important role in their formation and evolution. However, how baryonic matter is actually funneled into or ejected from star-forming regions in disk galaxies, like the Milky Way, is not clearly understood. 

In a grand picture of galaxy formation and evolution provided by $\Lambda$CDM cosmology, baryonic inflow into a galactic disk occurs when gas accretes from the inter-galactic medium into the dark matter halo and subsequently collapses toward its central region. However, past semi-analytic models, in which most of the gas in the halo are assumed to accrete into the galactic disk, predict that the slope at the faint end of the galaxy luminosity function is much steeper than that actually observed (e.g., White \& Frenk 1991). This implies that the gas cooling and accretion into the disk ought to be suppressed in real galaxies by feedback processes associated with, e.g., UV radiation, supernovae explosions and active galactic nuclei. These feedback processes are also necessary to reproduce the structural properties of disk galaxies. Indeed, earlier simulations with only modest feedback effects suffer from the so-called angular momentum problem that the simulated disk galaxies have too low angular momentum, thereby being more compact and denser than the observed ones (e.g., Navarro \& Steinmetz 2000). Recent high-resolution hydrodynamical simulations taking into account stronger feedback effects based on more realistic physical background have partially resolved such problems in galaxy formation (e.g., Okamoto et al. 2014; Vogelsberger et al. 2014; Sawala et al. 2014), but have not been completely successful yet (e.g., Sparre et al. 2014; Oman et al. 2015).

Feedback processes in galaxies are actually important as a driver of galactic outflow, which has been observed ubiquitously in starburst galaxies at both high and low redshifts (e.g., Shapley et al. 2003; Rupke et al. 2005; Tremonti et al. 2007; Weiner et al. 2009). It is especially remarkable from the observations of such galaxies that a large amount of metals are expelled from their galactic disks (e.g., Bouch\'e et al. 2005, 2006, 2007; Zahid et al. 2012; Peeples et al. 2014), thereby implying the importance of galactic outflow in the chemical evolution of both galaxies and inter-galactic medium.

To get more detailed insights into galactic inflow and outflow processes, it is worth investigating the global scaling relations of these star-forming galaxies, such as their stellar mass - metallicity (MZ) and stellar mass - star formation relations. For example, the MZ relation shows that at each redshift there is an empirical upper limit in metallicity for massive galaxies. This suggests that the chemical evolution in these galaxies accompanies gas inflow and outflow processes, unless otherwise the galactic metallicities, as derived in the closed-box models, are predicted to increase monotonously with increasing the stellar to gas mass ratio without any upper limits. Recently various studies based on numerical simulations and/or semi-analytic models have considered these global scaling relations to investigate the basic properties of gas inflow and outflow in normal star-forming galaxies (e.g., Dav\'e et al. 2011; Lilly et al. 2013; Kudritzki et al. 2015; Belfiore et al. 2015). 

We note that most of these previous studies set specific assumptions on the forms of gas inflow and outflow rates without explicit reasoning. For example, one of the popular assumptions is that inflow and outflow rates are always proportional to star formation rate (hereafter SFR) (e.g., Dayal et al. 2013; Yabe et al. 2015), which has been adopted in many studies without concrete assessment from the observational data. Therefore, in this paper, we investigate the general properties of gas inflow and outflow rates, in particular on their relation to SFR without prior assumptions. For this purpose, we focus on Milky Way-like galaxies (hereafter MWLGs), for which star formation histories are well investigated by the observational studies of distant galaxies. Provided that MWLGs follow the observed global scaling relations in their metallicities at all their evolutional stages, we obtain the inflow and outflow histories of MWLGs constrained by the observations of distant star-forming galaxies and derive their relation to SFR. 

In addition to these global relations provided by spatially unresolved observations, we also utilize the spatially resolved distributions of metals inside distant disk galaxies, which have been recently revealed by the observations with integrated field units (e.g., Cresci et al. 2010; Troncoso et al. 2014). Thus, combined with these data for internal metallicity distributions and their redshift dependence in distant disk galaxies, we derive the time evolution of gas inflow and outflow rates and their relation to SFR at each radius in MWLGs.

This paper is organized as follows. In Section 2, we present our method to derive the inflow and outflow rate densities from the observational results of distant star-forming galaxies. In Section 3, we show the inflow and outflow histories in MWLGs revealed from our analysis. In Section 4, we discuss the physical interpretations of the properties of the derived inflow and outflow rate densities. Finally, our conclusions are drawn in Section 5.

Throughout this paper, we use the following cosmological parameters: $H_0$ = 70 km s$^{-1}$ Mpc$^{-1}$, $\Omega_\Lambda$ = 0.7, and $\Omega_\mr{m}$ = 0.3, for all the relevant parts of the analysis.

%%% Sec.2 %%%%%%%%%%%%%%%%%%%%%%%%%%%%%%%%%%%%%%%%%%%%%%%%
\section{METHOD}
%%% Sec.2.1 %%%%%%%%%%%%%%%%%%%%%%%%%%%%%%%%%%%%%%%%%%%%%%%
\subsection{Basic Equations}
To investigate the surface densities of gas inflow and outflow rates in MWLGs at any time, $t$, and radius, $R$, along the galactic disk, denoted as $\sgmin \tr$ and $\sgmout \tr$, respectively, we solve the following equations (e.g., Pagel 1989; Chiappini et al. 1997),
\begin{equation}
\frac{\partial  \sgmg}{\partial t} = - (1 - \ret) \sgmsf + \sgmin - \sgmout \;,
\label{eq:dm1}
\end{equation}
\begin{equation}
\frac{\partial  (Z \sgmg)}{\partial t} = Y \sgmsf - Z (1 - \ret) \sgmsf + Z_\mr{in} \sgmin - Z_\mr{out} \sgmout \;.
\label{eq:dm2}
\end{equation}
Equations (\ref{eq:dm1}) and (\ref{eq:dm2}) represent the evolutions of a surface gas mass density, $\sgmg$, and of a mass fraction of heavy elements, $Z$, respectively. The first term on the right hand side of equation (\ref{eq:dm1}) describes the net gas consumption by star formation, where $\sgmsf$ and $\ret$ are the surface density of SFR and the mass fraction returned back to interstellar medium (ISM) via stellar mass loss, respectively, and the description of this term is based on an instantaneous recycling approximation. In this paper we set $\ret$ = 0.45 corresponding to the Chabrier initial mass function (Leitner \& Kravtsov 2011). This return fraction includes the contribution of stellar mass loss from low mass stars whose lifetime is as long as the age of the Universe. This may be somewhat problematic when studying an evolution of galaxy. However, according to Leitner \& Kravtsov (2011), the return fraction derived from stars whose lifetime is shorter than 1 Gyr is $\sim 0.38$, so that the mass returned back to ISM from low mass stars, for which an instantaneous recycling approximation may be invalid, is not significant. Therefore we expect that even if we adopt such a delayed stellar mass loss in our analysis, our conclusion in this paper remains basically unchanged. The second and third terms on the right hand side of equation (\ref{eq:dm1}) denote the contributions of inflow and outflow, respectively. The first term on the right hand side of equation (\ref{eq:dm2}) describes the supply of heavy elements newly synthesized in massive stars, where $Y$ is the nucleosynthetic yield, meaning the mass of heavy elements added into ISM per unit SFR. We assume that $Y$ is constant. In this paper we focus on the chemical abundance for oxygen, which has been estimated in many distant galaxies, and adopt the nucleosynthetic yield of $Y$ = 0.015 (Peeples et al. 2014, and references therein). The second term on the right hand side of equation (\ref{eq:dm2}) is the mass of heavy elements finally locked up in stars. The third and fourth terms on the right hand side of equation (\ref{eq:dm2}) represent the mass injection and ejection of heavy elements associated with inflow and outflow, respectively, where $Z_\mr{in}$ and $Z_\mr{out}$ are mass fractions of heavy elements in inflowing and outflowing gas, respectively. In these equations, we assume that inflowing gas contains no heavy elements and the metallicity of outflowing gas corresponds to that of the ISM, implying $Z_\mr{in} = 0$ and $Z_\mr{out} = Z$, similarly to most of the previous chemical evolution models (e.g., Dayal et al. 2013; Belfiore et al. 2015).

Using equations (\ref{eq:dm1}) and (\ref{eq:dm2}), we can calculate $\sgmin \tr$ and $\sgmout \tr$ if the three quantities, $\sgmg \tr$, $\sgmsf \tr$ and $Z \tr$, are given. To derive these three quantities in MWLGs, we adopt the recent observational results for distant star-forming galaxies as detailed in the next subsections.

These calculations are carried out in a radial range from $R$ = 0 to 10 kpc with a radial grid of $\Delta R$ = 0.1 kpc. This outer limiting radius roughly corresponds to the apertures of the high-redshift galaxies in their observations of SFR and chemical abundance (Mannucci et al. 2010, and references therein). Thus these observed quantities derived within the apertures of galaxies will be compared with those we derive within $R$ = 10 kpc in the model galaxy. For calculating the time evolution of the system, we adopt the time grid of $\Delta t$ = 100 Myr and confine ourselves to the redshift range of $z$ = 0 to 2, within which all the required observational information on the structural and chemical evolutions of MWLGs are available.

We note that equations (\ref{eq:dm1}) and (\ref{eq:dm2}) do not contain the effect of radial redistributions of heavy elements inside the galactic disk for the sake of simplicity in our following analysis. Such radial redistributions of metals can actually occur via radial gas flow associated with gravitational disk instabilities (e.g., Athanassoula 1992; Noguchi 1998) or re-accretion of metal-enriched gas ejected from the galactic disk as outflow (e.g., Oppenheimer \& Dav\'e 2008). We will discuss how these effects would modify our results in Section 4.4.

%%% Sec.2.2 %%%%%%%%%%%%%%%%%%%%%%%%%%%%%%%%%%%%%%%%%%%%%%%
\subsection{The Structural Evolution of MWLGs}
To estimate $\sgmg \tr$ and $\sgmsf \tr$ for MWLGs, we follow the results of van~Dokkum et al. (2013) (hereafter vD13), who investigated the stellar mass distribution of MWLG progenitors selected based on the abundance matching method from $z \sim$ 0 to 2.5. We adopt here the average evolution of stellar mass, $M_\ast$, S\' ersic index, $n$, and effective radius, $R_\mr{eff}$, of MWLGs obtained from the fittings to the results of vD13,
\begin{equation}
\mr{log}(M_\ast / \mr{M_\odot}) = 10.7 - 0.045 z - 0.13 z^2 \;,
\label{eq:ms}
\end{equation}
\begin{equation}
n = 2.74 - 0.147 z - 0.137 z^2 \;,
\label{eq:n}
\end{equation}
\begin{equation}
R_\mr{eff} = 3.42 - 0.182 z - 0.18 z^2 \;\; \mr{kpc} \;.
\label{eq:re}
\end{equation}
Equation (\ref{eq:ms}) implies that MWLGs had already obtained 80 \% of their present stellar mass by $z \sim$ 1. A similar mass evolution is expected from the analyses of disk stars in our Galaxy (Snaith et al. 2014). Equations (\ref{eq:n}) and (\ref{eq:re}) include both the bulge and disk components without decoupling each, so the S\' ersic index at $z$ = 0 is 2.74, which is greater than the typical value of a galactic disk with $n \sim 1$. The time evolution of a stellar mass profile led from equations (\ref{eq:n}) and (\ref{eq:re}) is such that prior to $z \sim 1$, the growth of stellar mass proceeds almost simultaneously at each radius within the disk, whereas, after $z \sim$ 1, the growth of stellar mass in the central region of the disk gradually decreases with time to make the stellar mass profile evolve in an inside-out manner. 

We here assume that all the stellar mass at each radius originates only from star formation, whereby there is no radial migration of stars in the galactic disk. This assumption allows us to derive $\sgmsf \tr$ simply from the stellar mass growth. We then translate $\sgmsf \tr$ into $\sgmg \tr$ based on the Kennicutt-Schmidt (KS) relation between SFR and surface gas density observed in local star-forming galaxies, i.e., $\sgmsf \propto \sgmg^\alpha$ with $\alpha \simeq$ 1.4 (Schmidt 1959; Kennicutt 1998). This assumption for star formation law may be reasonable because various studies have confirmed that the KS relation is basically satisfied up to $z \sim$ 2 (e.g., Genzel et al. 2010). We note here the recently updated knowledge that $\sgmsf$ cannot be simply represented by a single power law of $\sgmg$ but correlates rather with molecular gas density more tightly than atomic one (Bigiel et al. 2008). Gas in a galactic disk actually consists of various phases, such as neutral, ionized, molecular and dust components, and each state can be associated with star formation in a different way. However in this work, we simply define the gas density as basically the sum of atomic and molecular gas densities to avoid the complexity of the transformation between these different states.

With the above procedure, we are thus able to obtain both $\sgmg \tr$ and $\sgmsf \tr$ of MWLGs from the results of vD13. Recently, based on the similar method to vD13, Morishita et al. (2015) also investigated the stellar mass distribution of MWLG progenitors from $z \sim$ 0.5 to 3, but adopting a different stacking analysis, which takes into account an position angle and axis ratio of each MWLG progenitor. Their results partly disagree with vD13 in the point that the mass in the central region of the MWLG progenitor increases with time even after $z \sim$ 1. We use their results to calculate $\sgmg \tr$ and $\sgmsf \tr$ and find that the derived histories of gas inflow and outflow of MWLGs do not significantly depend on the difference in the adopted evolution of the stellar mass profile. In what follows, we present only the results derived from the results of vD13.

We note here that the assumption of no radial redistribution of stars in the galactic disk may be too simplistic, because various theoretical and observational studies have suggested the importance of radial migration of disk stars (e.g., Sellwood \& Binney 2002; Kordopatis et al. 2015; Morishita et al. 2015; Hayden et al. 2015). However, how actually the evolution of a stellar mass profile depends on the radial migration process is not clearly understood yet. Some specific discussion on the effect of radial migration in our analysis may be possible by adopting chemo-dynamical models such as in Sch\" onrich \& Binney (2009) and Kubryk et al. (2014). However quantifying the effect firmly is yet difficult because the predicted properties of radial migration are significantly different between models. Although the subject is important and worth investigating, we attempt to keep the clarity of the following analysis by neglecting this radial migration effect and thus avoiding the associated uncertainties.

%%% Sec.2.3 %%%%%%%%%%%%%%%%%%%%%%%%%%%%%%%%%%%%%%%%%%%%%%%
\subsection{The Chemical Evolution of MWLGs}
How the chemical evolution in MWLGs proceeds can be inferred from the metallicity measurements for distant star-forming galaxies. Mannucci et al. (2010) showed that at $z \lesssim 2.5$ the average chemical abundances of such galaxies can be described as the following function of their stellar mass and SFR,
\begin{eqnarray}
\begin{split}
12 + \mr{log(O/H)} = 8.90 &+ 0.39x - 0.20x^2 \\
&- 0.077x^3 + 0.064x^4 \;,
\label{eq:fmr}
\end{split}
\end{eqnarray}
where $x = \mr{log}(M_\ast/\mr{M_\odot}) - 0.32 \mr{log}(\mr{SFR}/\mr{M_\odot\;yr^{-1}}) - 10$. This tight relation, called the fundamental metallicity relation (hereafter FMR), implies that most of star-forming galaxies follows the FMR at all epochs since $z \sim 2.5$. In this work, we regard the metallicity derived from the FMR as the average metallicity of gas within the galaxy. However, it is worth noting that this may not necessarily be a proper approximation, because the metallicity based on an analysis of line emissions preferentially reflects metallicities of gas in star forming regions, where most of photons of useful emission lines in the analysis, originate. The relation between metallicity derived from integrated spectrum within aperture and that properly averaged in the galaxy will be an important subject in the future studies with both spatially resolved and unresolved data of galaxies.

Combining the FMR with the stellar mass evolution of vD13 can provide the total chemical abundance and its time evolution in MWLGs, $Z_\mr{tot}(t)$. We define $Z_\mr{tot} = \int^{R_\mr{max}}_0 Z \sgmg R \mr{d}R / \int^{R_\mr{max}}_0 \sgmg R \mr{d}R$ with $R_\mr{max}$ = 10 kpc, which suggests that $Z \tr$ can be calculated from the evolution of the radial metallicity gradient (hereafter RMG), $\mr{d}Z\tr/\mr{d}R$, along the disk. The RMGs of distant star-forming galaxies have been derived by dedicated observations with integral field units such as VLT/SINFONI and VLT/KMOS. Stott et al. (2014) showed that the RMGs measured for star-forming galaxies at $z \lesssim 1.5$ can be described as an increasing function with increasing specific star formation rate (sSFR) as follows, 
\begin{eqnarray}
\begin{split}
\frac{\Delta \mr{log(O/H)}}{\Delta R} = \;&0.023 \; \mr{log(sSFR/yr^{-1})} \\
&+ 0.20 \;\; \mr{dex\;kpc^{-1}}\;,
\label{eq:rmg}
\end{split}
\end{eqnarray}
namely the slope of the radial metallicity distribution becomes steeper with decreasing sSFR. This suggests that if MWLGs follow this relation, their RMG gradually steepens with time because sSFR is generally higher at higher redshifts. Such evolution of the RMG is also shown in Gibson et al. (2013), who simulated the chemical evolution of disk galaxies with strong feedback. Also, the RMGs of older disk stars in our Galaxy are found to be flatter or positive (Casagrande et al. 2011; Toyouchi \& Chiba 2014). 

%%% Figure 1 %%%

\begin{figure}
\figurenum{1}
\begin{center}
\includegraphics[width=8.0cm,height=5.0cm]{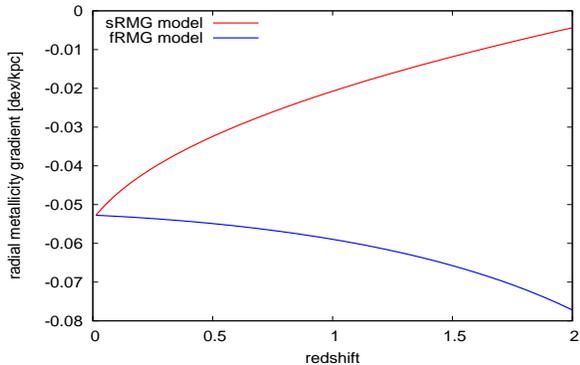}
\end{center}
\caption{The redshift evolutions of the radial metallicity gradients for the sRMG (red line) and fRMG (blue line) models.}
\label{fig:rmg} 
\end{figure}

The red line in Figure \ref{fig:rmg} shows the evolution of the RMG derived by combining the RMG vs. sSFR relation in Stott et al. (2014) with the stellar mass evolution in equation (\ref{eq:ms}). The derived RMG at $z = 0$ is $-0.053$ $\mr{dex\:kpc^{-1}}$, which is slightly flat compared to the present RMG in our Galaxy of $\Delta \mr{[Fe/H]} / \Delta R = -0.062 \;\mr{dex\;kpc^{-1}}$ (Luck \& Lambert 2011). We adopt this steepening RMG (hereafter sRMG) with time as a standard case to derive $Z \tr$, with which we investigate gas inflow and outflow histories in MWLGs.

Although this sRMG model is actually inferred from observations, we also consider the different time evolution of the RMG in a manner that it is more flattened with time, which may be the case if disk galaxies form in an inside-out process. This model is referred to as the flattening RMG (hereafter fRMG) model in comparison with the sRMG model. To construct this model, we make use of the observational fact that the RMGs observed in local star-forming galaxies are proportional to $R_\mr{eff}^{-1}$ (e.g., Vila-Costas \& Edmunds 1992; S\' anchez et al. 2014), although the observations of high-$z$ galaxies show an opposite result (Stott et al. 2014). If we adopt this relation between the RMG and effective radius, the RMG gradually flattens with time because the effective radius of MWLGs is an increasing function of time. 

The blue line in Figure \ref{fig:rmg} represents the fRMG model derived by combining the relation, $\mr{RMG} = \mr{RMG}_0 \; (R_\mr{eff}(z)/R_\mr{eff}(z=0))^{-1}$, and equation (\ref{eq:re}), where $\mr{RMG}_0$ is the RMG at $z$ = 0, and is set as the same as that for the sRMG model. The comparison between these two models for the time evolution of the RMGs allows us to investigate how the difference in it affects the estimate of inflow and outflow rates in MWLGs.

%%% Sec.3 %%%%%%%%%%%%%%%%%%%%%%%%%%%%%%%%%%%%%%%%%%%%%%%%
\section{RESULTS}
%%% Sec.3.1 %%%%%%%%%%%%%%%%%%%%%%%%%%%%%%%%%%%%%%%%%%%%%%%%
\subsection{sRMG Model}
\begin{figure}
\figurenum{2}
\begin{center}
\includegraphics[width=8.0cm,height=10.0cm]{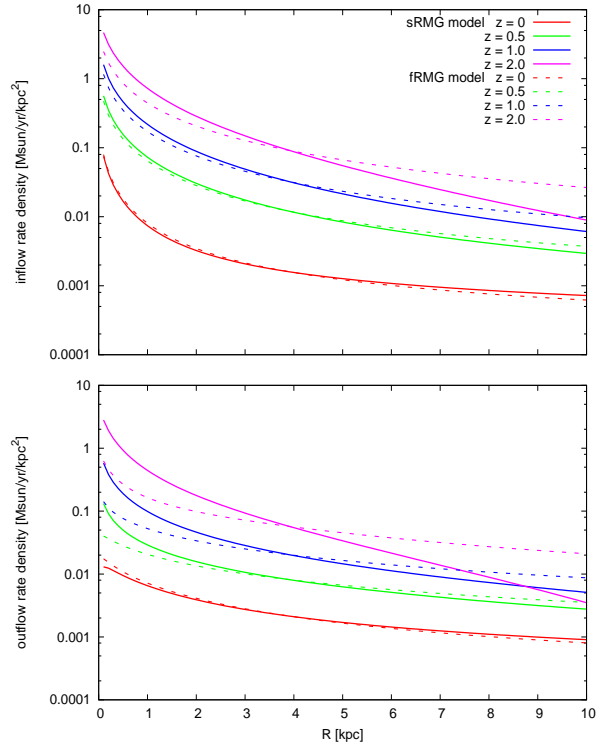}
\end{center}
\caption{The inflow and outflow rate densities (top and bottom panels, respectively) as a function of the distance from the galactic center at redshifts of $z =$ 0 (red), 0.5 (green), 1 (blue) and 2 (purple). The results for the sRMG and fRMG models are shown with solid and dashed lines, respectively.}
\label{fig:1} 
\end{figure}

We obtain the inflow and outflow rate densities, $\sgmin$ and $\sgmout$, by substituting $\sgmg$, $\sgmsf$ and $Z$ derived in Section 2.2 and 2.3 into equations (\ref{eq:dm1}) and (\ref{eq:dm2}). Here we present the results for the sRMG model. Solid lines in Figure \ref{fig:1} show $\sgmin$ (top panel) and $\sgmout$ (bottom panel) as a function of $R$ at redshifts of 0, 0.5, 1.0 and 2.0. It is clear that both $\sgmin$ and $\sgmout$ are higher in the inner disk region at all epochs and that their central values monotonically decline with decreasing redshifts. We also find that the radial profiles of $\sgmin$ and $\sgmout$ evolve roughly in an inside-out manner rather than a simultaneous one, in contrast to the evolution of the stellar mass profile introduced in Section 2.2.

%%% Figure 2 %%%

%%% Figure 3 %%%

\begin{figure}
\figurenum{3}
\begin{center}
\includegraphics[width=8.0cm,height=10.0cm]{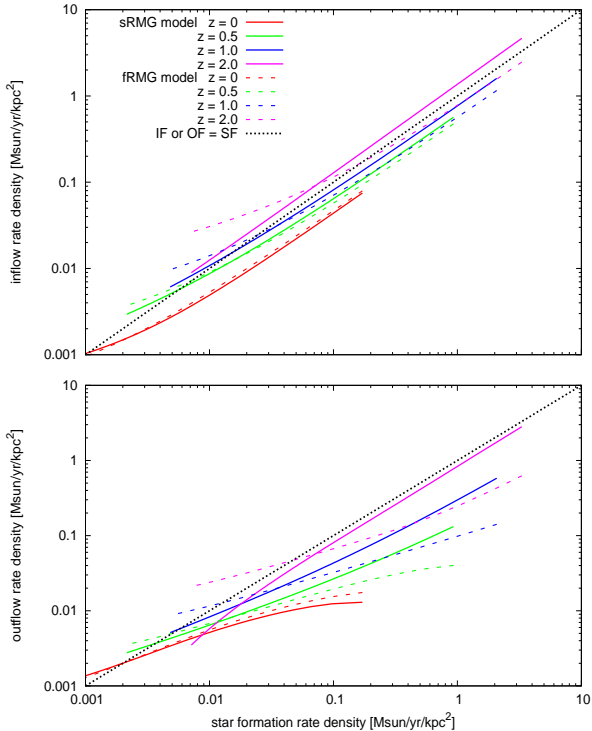}
\end{center}
\caption{$\sgmin$ vs. $\sgmsf$ (top panel) and $\sgmout$ vs. $\sgmsf$ (bottom panel) at redshifts of $z =$ 0 (red), 0.5 (green), 1 (blue) and 2 (purple). The results for the sRMG and fRMG models are shown with solid and dashed lines, respectively.}
\label{fig:2} 
\end{figure}

To get further insights into the properties of gas inflow and outflow in MWLGs, we here investigate the dependences of $\sgmin \tr$ and $\sgmout \tr$ on $\sgmsf \tr$. This procedure allows us to inspect whether the assumption adopted in many previous studies, namely that inflow and outflow rates in galaxies are proportional to their SFR, is indeed valid or not. Solid lines in Figure \ref{fig:2} show $\sgmin$ and $\sgmout$ for the sRMG model as a function of $\sgmsf$ at redshifts of 0, 0.5, 1.0 and 2.0. Black dotted lines show $\sgmin = \sgmsf$ (top panel) and $\sgmout = \sgmsf$ (bottom panel) for comparison. It follows that the inflow rate density roughly satisfies a proportionality to the SFR density at all epochs since $z = 2$. For outflow, while $\sgmout$ is roughly proportional  to $\sgmsf$ at high $z$, this proportionality appears broken at low $z$.

\begin{figure}
\figurenum{4}
\begin{center}
\includegraphics[width=8.0cm,height=10.0cm]{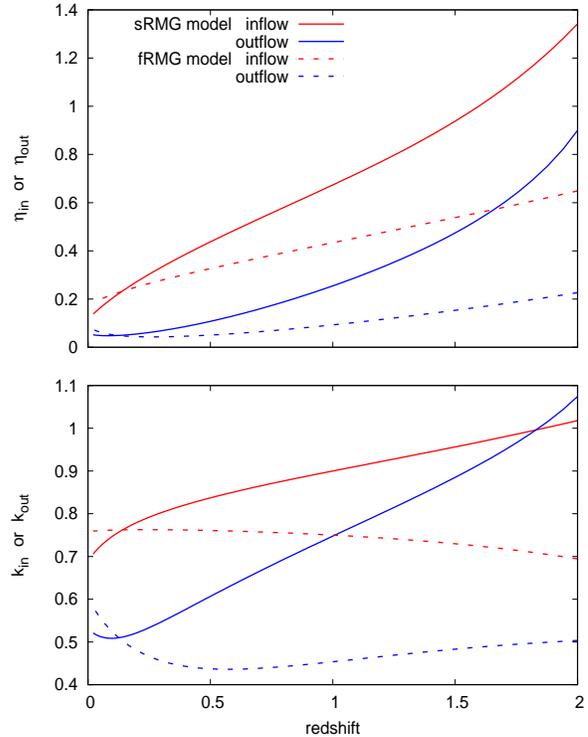}
\end{center}
\caption{The redshift evolutions of the parameters, $\eta_\mr{in/out}$ (top panel) and $k_\mr{in/out}$ (bottom panel), defined in equation (\ref{eq:fit}). The results for gas inflow and outflow are shown with red and blue lines, respectively, where the solid and dashed lines correspond to the results for the sRMG and fRMG models, respectively.}
\label{fig:3} 
\end{figure}

To describe these redshift evolutions of the relations between the inflow/outflow and SFR densities in a more explicit form, we fit $\sgmin$ and $\sgmout$ at each epoch to the following power-law functions of $\sgmsf$,
\begin{equation}
\sgmin =  \etain \sgmsf^{\kin} \;,\;\;\; \sgmout =  \etaout \sgmsf^{\kout}  \;.
\label{eq:fit}
\end{equation}
These parameters, $\eta_\mr{in/out}$ and $k_\mr{in/out}$, correspond to the intercepts and slopes of the profiles shown in Figure \ref{fig:2}, respectively. In the top panel of Figure \ref{fig:3}, we show the redshift evolutions of $\etain$ and $\etaout$ with red and blue solid lines, respectively, indicating that $\etain$ and $\etaout$ increase with increasing redshift. Also, $\etain$ is always larger than $\etaout$, and is larger than unity at $z \gtrsim 1.5$, so that the inflow rate dominates the SFR at this early epoch. The bottom panel in Figure \ref{fig:3} shows the redshift evolution of $\kin$ and $\kout$ with red and blue solid lines, respectively. We find that $\kin$ decreases with time, but is nearly unity at all epochs, implying the proportional relation between the inflow rate and SFR as mentioned above. In contrast to the evolution of $\kin$, $\kout$ evolves from $\sim$ 1.0 at $z = 2$ to $\sim$ 0.5 at $z = 0$. This result suggests that the proportionality of the outflow rate to the SFR as adopted in some of previous galactic evolution models is not necessarily valid.

%%% Figure 4 %%%

%%% Sec.3.2 %%%%%%%%%%%%%%%%%%%%%%%%%%%%%%%%%%%%%%%%%%%%%%%%
\subsection{fRMG Model}
The results for this model are shown with dashed lines in Figures \ref{fig:1}, \ref{fig:2} and \ref{fig:3} in comparison to the sRMG model with solid lines. We find that although the general evolution of $\sgmin$ and $\sgmout$ for the fRMG model are similar to those for the sRMG model, there is a remarkable difference between the two models that the radial dependence of $\sgmin$ and $\sgmout$ for the fRMG model shows no significant evolution and remains simultaneous at each epoch, in contrast to the inside-out evolution of $\sgmin$ and $\sgmout$ for the sRMG model.

These common or different properties of $\sgmin$ and $\sgmout$ between the two models can be understood in terms of the redshift evolution of $\eta_\mr{in/out}$ and $k_\mr{in/out}$. The top panel of Figure \ref{fig:3} shows that for the fRMG model both $\etain$ and $\etaout$ decrease with time and $\etain$ is always larger than $\etaout$. These properties of $\etain$ and $\etaout$ for this model are the same as those for the sRMG model, implying that such properties may reflect the evolution of the total stellar mass or entire chemical abundance in MWLGs rather than the evolution of the RMG in the galactic disk. In fact, a similar result was reported by Yabe et al. (2015), who investigated the general properties of galactic inflow and outflow by analyzing the scaling relations of star-forming galaxies between stellar mass, gas mass fraction and chemical abundance at each redshift. However, there is a difference between the results of the two models that $\etain$ and $\etaout$ for the fRMG model are much smaller than those for the sRMG model at high $z$. This may be because for the fRMG model the suppression effect of chemical evolution due to gas inflow and outflow is smaller than that for the sRMG model to make the chemical abundance of the central regions of the disk remain higher at earlier epochs, so that the RMG is made flatter at subsequent epochs. 

From the bottom panel in Figure \ref{fig:3}, we can confirm the proportionality of the inflow rate to SFR in the fRMG model, i.e., $\sgmin \propto \sgmsf^{\kin}$ with $\kin \sim 1$, as also seen in the sRMG model. The outflow rate is roughly proportional to the square root of the SFR at all redshifts, $\sgmout \propto \sgmsf^{\kout}$ with $\kout \sim 1/2$, in contrast to the sRMG model. This difference in the redshift evolution of $\kout$ between two models reflects the difference in the adopted RMG evolution, which will be discussed in detail in Section 4.1.

%%% Sec.4 %%%%%%%%%%%%%%%%%%%%%%%%%%%%%%%%%%%%%%%%%%%%%%%%
\section{DISCUSSION}
As presented in the previous sections, we parameterize the gas inflow and outflow rates in MWLGs in terms of $\eta_\mr{in/out}$ and $k_\mr{in/out}$ as described in equation (\ref{eq:fit}).  The properties of these parameters are summarized as follows.
\begin{itemize}
\item Both $\etain$ and $\etaout$ are a decreasing function of time.
\item The index $\kin$ is almost unity at all epochs, implying that the inflow rate can be regarded to be proportional to the SFR.
\item The index $\kout$ for the fRMG model is about 0.5 at all epochs, whereas that for the sRMG model evolves from $\sim$ 1 at $z = 2$ to $\sim$ 0.5 at $z = 0$. Therefore the evolution of $\kout$ is expected to depend on the adopted evolution of the RMG.
\end{itemize}
These results reflect basic baryonic physics associated with gas inflow and outflow processes in the disk. In this section, we discuss how the above properties of $\eta_\mr{in/out}$ and $k_\mr{in/out}$ are related to and deduced from such baryonic physics.

%%% Sec.4.1 %%%%%%%%%%%%%%%%%%%%%%%%%%%%%%%%%%%%%%%%%%%%%%%%
\subsection{The Relation between the Evolutions of $\kout$ and the RMG}
The most remarkable difference between the results of the sRMG and fRMG models is the redshift evolution of $\kout$. This difference suggests that the dependence of outflow process on star formation activity affects the chemical evolution over the galactic disk significantly. The relation between $\kout$ and the RMG can be simply interpreted as follows. 

For an ordinary galactic disk that has more active star formation in its inner region, $\kout > 1$ means that the ratio of the outflow rate to SFR becomes higher toward the inner region. Therefore, the enrichment process driven by active star formation in the inner region of the disk is suppressed by the effect of much higher outflow rate of gas, so that the steepening of the RMG with time is suppressed, although the actual situation is expected to depend on the nucleosynthetic yield and initial distribution of heavy elements in the disk. In contrast, for $\kout < 1$ the chemical evolution always progresses faster in the inner disk region, so that the RMG is steepened with time. Therefore, for the sRMG model $\kout \gtrsim 1$ at high $z$ leads to the flat RMG and the gradual decrease of $\kout$ accompanies the steepening of the RMG. On the other hand, for the fRMG model $\kout \ll 1$ at all redshifts makes the RMG steeper at earlier epochs. 

Thus the chemical evolution over the galactic disk strongly depends on the properties of the outflow process, whereby the physical origin of the evolution of $\kout$ is a key to understanding the chemical evolution of the galactic disk. In the next subsection, we present possible models for explaining these properties of the outflow process.

%%% Sec.4.2 %%%%%%%%%%%%%%%%%%%%%%%%%%%%%%%%%%%%%%%%%%%%%%%%
\subsection{The Relation between $\sgmout$ and $\sgmsf$}
Previous studies with both analytical models and numerical simulations have suggested that gas outflows from a galactic disk are driven by the injection of kinetic energy (e.g., Springel \& Hernquist 2003; Okamoto et al. 2010) or momentum (e.g., Murray et al. 2005; Oppenheimer \& Dav\'e 2006; Hopkins et al. 2012) from stellar feedback into ISM. The former and latter outflow mechanisms are called the energy-driven wind (EDW) and momentum-driven wind (MDW), respectively. Recently Okamoto et al. (2014) showed that a baryonic feedback model including both EDW and MDW successfully reproduces the various properties of galaxies, for example stellar mass function, stellar mass - halo mass relation and their redshift evolutions from $z$ = 4 to 0. However, which wind mechanism is actually the main source of galactic outflow is not well understood. In this section we consider these EDW and MDW models as the origin of the relation between $\sgmout$ and $\sgmsf$ obtained in our analysis, which will provide valuable insights into the main mechanism of galactic outflow.

%%% Sec.4.2.1 %%%%%%%%%%%%%%%%%%%%%%%%%%%%%%%%%%%%%%%%%%%%%%%%
\subsubsection{Energy-driven wind}
First, we attempt to describe $\sgmout$ as a function of $\sgmsf$ in the framework of the EDW model. Here we consider a star cluster with stellar mass of $\ms$. Because the energy injected from the star cluster into ISM is proportional to $\ms$ in the EDW model, the mass of outflow ejected from the star cluster, $\mw$, is described by using a typical velocity of outflowing gas when leaving the disk, $\vw$, as follows,
\begin{equation}
\mw \propto \frac{\ms}{\vw^2} \;.
\label{eq:edw1}
\end{equation}
We assume that $\vw$ corresponds to the escape velocity of the disk defined as,
\begin{equation}
\vw \sim \sqrt{\pi G H (\sgms + \sgmg)} \propto \sqrt{(1+\mu^{-1})\sgmg} \;,
\label{eq:vw}
\end{equation}
where $\mu \equiv \sgmg/\sgms$, and $H$ is the scale height of the disk and is assumed to be constant along the radius. As the number of star clusters with $\ms$ in the region where stars are formed in the rate of $\sgmsf$ is estimated as $\sgmsf / \ms$, we can describe $\sgmout$ in terms of $\sgmsf$ as follows,
\begin{equation}
\sgmout \sim \mw \frac{\sgmsf}{\ms} \;.
\label{eq:out}
\end{equation}
By substituting equations (\ref{eq:edw1}) and (\ref{eq:vw}) into (\ref{eq:out}) and adopting the KS relation, $\sgmsf \propto \sgmg^\alpha$, we express $\sgmout$ as functions of $\mu$ and $\sgmsf$ as follows,
\begin{equation}
\sgmout \propto \frac{\mu}{1+\mu} \; \sgmsf^{(\alpha-1)/\alpha} \;.
\label{eq:edw2}
\end{equation}
We regard the factor $\mu/(1+\mu)$ in this equation depends only on time, because the variation of $\mu$ along the radius is much smaller than that of $\sgmsf$. Therefore $\mu/(1+\mu)$ and $(\alpha-1)/\alpha$ in this equation correspond to $\etaout$ and $\kout$ in equation (\ref{eq:fit}), respectively. 

\begin{figure}
\figurenum{5}
\begin{center}
\includegraphics[width=8.0cm,height=5.0cm]{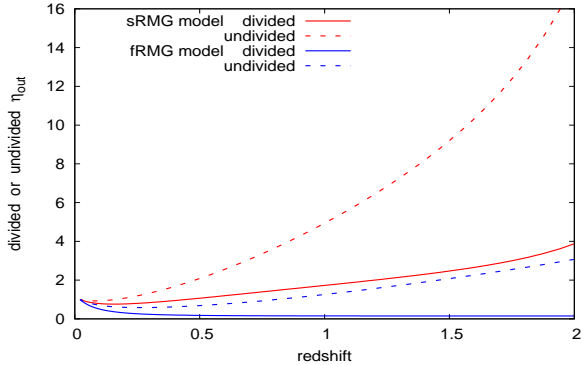}
\end{center}
\caption{The redshift evolution of $\etaout$ divided by $\sqrt{\mu/(1+\mu)}$ and $\mu/(1+\mu)$ for the sRMG model (red solid line) and fRMG model (blue solid line), respectively. The parameter $\mu$ used here is calculated from the total gas and stellar mass at each redshift. For comparison, we also show the undivided $\etaout$ for the sRMG model (red dashed line) and fRMG model (blue dashed line). Each line is normalized by the value at $z$ = 0.}
\label{fig:e/m} 
\end{figure}

This formulation reveals that $\kout$ is independent of time and that by adopting $\alpha = 1.4$ as observed in local star-forming galaxies we find $\kout = (\alpha-1)/\alpha \sim 0.29$. These properties of $\kout$ are basically in agreement with our results obtained for the fRMG model. Moreover, since $\mu$ is generally much higher at higher redshifts, we reproduce $\etaout$ as a decreasing function of time. To confirm the validity of equation (\ref{eq:edw2}), in Figure \ref{fig:e/m} we show $\etaout$ divided by the factor $\mu/(1+\mu)$ with blue solid line for the fRMG model. In this figure we also show the undivided $\etaout$ with blue dashed line for comparison. We note that each $\etaout$ is normalized by the value at $z$ = 0, and $\mu$ used in this procedure is derived by using the total gas and stellar mass at each redshift. We find that the divided $\etaout$ for the fRMG model is only slightly smaller than unity at all redshifts and is approximately constant at $z \gtrsim 0.25$. Thus the EDW model can well represent the properties of gas outflow obtained for the fRMG model. In contrast, the properties of $\kout$ in the sRMG model are not understood in the framework of the EDW model but the MDW model is at work as explained below.

%%% Figure 5 %%%

%%% Sec.4.2.2 %%%%%%%%%%%%%%%%%%%%%%%%%%%%%%%%%%%%%%%%%%%%%%%%
\subsubsection{Momentum-driven wind}
We now derive the relation between $\sgmout$ and $\sgmsf$ based on the MDW model. We consider that the radiation emitted from a star cluster drives the ambient gas as outflow. The momentum injected into the ambient gas from a star cluster via the radiation during the time scale of $\tau_\mr{inj}$, which should roughly correspond to the lifetime of massive stars, i.e., $\sim$ 10 Myr, is described as follows,
\begin{equation}
P \sim (1+\tau_\mr{IR}) \; \frac{L}{c} \;\tau_\mr{inj} \;,
\label{eq:dp}
\end{equation}
where $L$ and $\tau_\mr{IR}$ are the luminosity of the star cluster and the optical depth of the ambient gas to the infrared emission by dust grains, respectively. 

We note that $\tau_\mr{IR}$ is proportional to the surface gas density of giant molecular clouds (GMCs) surrounding the star cluster, $\Sigma_\mr{GMC}$, rather than $\sgmg$ as in equations (\ref{eq:dm1}) and (\ref{eq:dm2}). Therefore, in order to discuss the result in Figure \ref{fig:3}, we need to explicitly describe the relation between $\Sigma_\mr{GMC}$ and $\sgmg$. We consider here that each of a GMC is originally a fragment of the galactic gas disk with $\sgmg$ and the scale of fragmentation is roughly consistent with the scale height of the gas disk. Then the mass of a GMC can be described as $M_\mr{GMC} = \pi \sgmg H^2 = \pi \Sigma_\mr{GMC} r_\mr{GMC}^2$, where $r_\mr{GMC}$ is the radius of a GMC. Assuming that the ratio of $H$ to $r_\mr{GMC}$ is nearly constant, as in Murray et al. (2011), we obtain $\tau_\mr{IR} \propto \sgmg$. In this manner, we connect the physics in a galactic disk scale with that in a GMC scale in this discussion.

Here we consider the optically thin ($\tau_\mr{IR} \ll 1$) and thick ($\tau_\mr{IR} \gg 1$) limits, and the mass of outflow ejected from the star cluster for each limit is described as follows,
\begin{eqnarray}
\begin{split}
\mw \sim
	\begin{cases}
		\tau_\mr{inj} (L/c)/\vw \propto \ms/\vw \\
		\;\;\;\;\;\;\;\;\;\;\;\;\;\;\;\;\;\;\;\;\;\;\; \text{(optically thin limit)} \\
		\tau_\mr{inj} (\tau_\mr{IR}L/c)/\vw \propto \sgmg \; \ms/\vw \\
		\;\;\;\;\;\;\;\;\;\;\;\;\;\;\;\;\;\;\;\;\;\;\; \text{(optically thick limit)} \;,
	\end{cases}
\label{eq:mdw1}
\end{split}
\end{eqnarray}
where we assume that $L$ is proportional to $\ms$. From equations (\ref{eq:vw}), (\ref{eq:out}) and (\ref{eq:mdw1}), $\sgmout$ for the MDW model can be presented as follows;
\begin{eqnarray}
\sgmout \propto 
	\begin{cases}
		\sqrt{\frac{\mu}{1+\mu}} \; \sgmsf^{(2\alpha-1)/2\alpha} & \text{(optically thin limit)} \\
		\sqrt{\frac{\mu}{1+\mu}} \; \sgmsf^{(2\alpha+1)/2\alpha} & \text{(optically thick limit)} \;.
	\end{cases}
\label{eq:mdw2}
\end{eqnarray}
Thus, we find that $\sgmout$ for the MDW model can also be expressed in the same functional form as equation (\ref{eq:fit}), similarly to the EDW model. 

The remarkable difference from the case of the EDW model is that $\kout$ depends on the optical depth of the ambient gas and thus the surface gas density. Adopting $\alpha = 1.4$ leads to $\kout \sim$ 0.64 and 1.36 for the optically thin and thick limit, respectively, which roughly reproduce the values of $\kout$ for the sRMG model at $z \sim$ 0 and 2, respectively. Therefore, we suggest that the evolution of $\kout$ obtained for the sRMG model is understood in the framework of the MDW model, as the transition from the optically thick to thin limit in the galactic gas disk. This transition occurs at the gas density of $\sim 100 \;\mr{M_\odot} \; \mr{pc^{-2}}$ (Murray et al. 2011), which would be sufficiently lower than the gas densities in high-$z$ MWLGs. 

Equation (\ref{eq:mdw2}) also predicts the decreasing $\etaout$ with time. The red solid and dashed lines in Figure \ref{fig:e/m} show $\etaout$ for the sRMG model divided and undivided by the factor $\sqrt{\mu/(1+\mu)}$, respectively. From this figure, we find that the divided $\etaout$ remains nearly unity and constant in comparison with the undivided $\etaout$. Thus MDW appears to explain well the properties of gas outflow in the sRMG model. 

From the above discussion, we conclude that clarifying how the time evolution of the RMG in MWLGs proceeds, i.e., more steepened or flattened with time, distinguishes which wind-blowing mechanism, injection of momentum or kinetic energy from stellar feedback, is actually at work. Thus, future observational constraints on the RMG in MWLGs will be important to understand the main mechanism driving galactic outflows. 

%%% Sec.4.3 %%%%%%%%%%%%%%%%%%%%%%%%%%%%%%%%%%%%%%%%%%%%%%%%
\subsection{The Proportionality of $\sgmin$ to $\sgmsf$}
In Section 3, we have shown the linear relation between $\sgmin$ and $\sgmsf$, which is almost independent of the difference in the RMG evolutions. A similar proportionality was also confirmed by  Recchi et al. (2008), who compared a chemical evolution model including gas inflow being proportional to SFR with a model assuming exponentially decreasing inflow rate with time. We show here that this proportionality is indeed universal for all the galactic evolution models with various gas inflow histories. 

Following the results of this work as well as previous works, gas inflow is always systematically larger than outflow, thereby we assume here $\sgmin - \sgmout \sim \sgmin$ as a working hypothesis in the following discussion. Then equation (\ref{eq:dm1}) for the conservation of gas mass is rewritten as,
\begin{equation}
\sgmin = \frac{\mr{d}\sgmg}{\mr{d} t} + (1 - \ret) \sgmsf \;.
\label{eq:sgmin1}
\end{equation}
Since $\mr{d}\sgmg/\mr{d}t$ is described with $\mu$ as follows,
\begin{equation}
\frac{\mr{d}\sgmg}{\mr{d} t} = \sgmg \frac{\mr{d} \mr{ln} \mu}{\mr{d} t} + \mu (1-\ret)\sgmsf \;,
\label{eq:sgmin2}
\end{equation}
then, by substituting this equation into equation (\ref{eq:sgmin1}), $\sgmin$ can be expressed as,
\begin{equation}
\sgmin = \left \{ (1+\mu) (1-\ret) + \frac{\sgmg}{\sgmsf} \frac{\mr{d} \mr{ln} \mu}{\mr{d} t} \right \} \sgmsf \;.
\label{eq:sgmin3}
\end{equation}
The second term in the bracket on the right hand side corresponds to the ratio of gas consumption timescale, $\tau_\mr{gas}$ ($\equiv \sgmg/\sgmsf$), to the timescale on which $\mu$ is changing, $\tau_\mu$ ($\equiv \mr{d}t/\mr{d ln}\mu$). We note here that $\tau_\mu$ depends on both SFR and inflow rate, whereas $\tau_\mr{gas}$ depends only on SFR. This implies that while $\tau_\mr{gas} \sim \tau_\mu$ in the closed-box model, the model taking into account inflow yields $\tau_\mr{gas} < \tau_\mu$, because the gas supply by inflow cancels out the decrease of the gas mass due to star formation. Therefore we find from equation (\ref{eq:sgmin3}) that $\sgmin$ is basically proportional to $\sgmsf$ for $\mu \gtrsim 1$, which is in particular the case at high redshifts. Then the property of $\etain$ is dominated by the factor $(1+\mu)(1-\ret)$, which is a decreasing function of time, thereby reproducing the time evolution of $\etain$ presented in Section 3. For $\mu < 1$, as expected at most of the lower redshifts, the second term in the bracket on the right hand side of equation (\ref{eq:sgmin3}) provides the deviation from the simple proportionality, which may reproduce $\kin$ being slightly smaller than unity as derived in our analysis. 

Thus the proportionality of $\sgmin$ to $\sgmsf$ as revealed by our analysis is simply understood from the conservation of gas mass. This proportionality may imply that gas inflow regularly controls the gas budget and SFR in the galactic disk.

%%% Sec.4.4 %%%%%%%%%%%%%%%%%%%%%%%%%%%%%%%%%%%%%%%%%%%%%%%%
\subsection{The Influence of Radial Redistributions of Metals on Our Results}
As noted in Section 2.1, our analysis does not include the effect of radial redistributions of metals inside the galactic disk. However, the recent observations of nearby galaxies having extended gas disks show that the star formation activities observed at their outskirts are too weak to produce their metallicities observed in such regions, and therefore suggest the need of significant transportations of heavy elements from their inner to outer disk regions (Werk et al. 2010, 2011; Bresolin et al. 2012). Such transportations of heavy elements may be driven by radial flows of gas inside galactic disks due to gravitational instabilities or re-accretion of metal-enriched galactic wind into the galactic disk. In this subsection we discuss the possible impacts of such radial redistributions of heavy elements on our results.

%%% Sec.4.4.1 %%%%%%%%%%%%%%%%%%%%%%%%%%%%%%%%%%%%%%%%%%%%%%%%
\subsubsection{Radial flows of gas}
Radial flow of gas may be frequently driven by non-axisymmetric structures such as bar/spiral and significantly affect the structural and chemical evolutions of disk galaxies (e.g., Sch\"onrich \& Binney 2009; Minchev et al. 2013; Kubryk et al. 2014). We here attempt to simply evaluate to what extent such radial flows can actually change our results. It is worth noting, however, that the answer may sensitively depend on the detailed properties of given radial flow.

The effect of radial flow on the evolution of a surface gas mass density at each radius is represented by adding an advection term to the right side of equation (1), $-\partial (R \sgmg v_R) / R \partial R$, where $v_R$ is the velocity of radial flow. Taking into account $\sgmin \gg \sgmout$ as obtained above, we write 
\begin{equation}
\frac{\partial  \sgmg}{\partial t} \sim - (1 - \ret) \sgmsf + \sgmin - \frac{1}{R} \frac{\partial}{\partial R} \left ( R \sgmg v_R \right)  \;.
\label{eq:rf}
\end{equation}
Several chemical evolution models for our Galaxy including radial flow suggested the advection term in equation (\ref{eq:rf}) to be of the order of $\sim 0.01 M_\odot\;\mr{yr^{-1}\;kpc^{-2}}$ (e.g., Sch\"onrich \& Binney 2009; Kubryk et al. 2014), which is comparable to $\sgmin$ at $z \lesssim 0.5$ as followed from Figure \ref{fig:1}. Therefore the influence of radial flow on the estimate of $\sgmin$ may be significant only at $z \lesssim 0.5$. We also note that in our analysis all the terms except for those of $\sgmin$ and advection in equation (\ref{eq:rf}) are derived independently from the latter two terms, so that the change of $\sgmin$ associated with radial flow simply reflects the sign and amplitude of the advection term. 

We now consider the effect of bar/spiral structures, which generally drive radial flow with negative (positive) $v_R$ for the gas at $R$ smaller (larger) than the co-rotation radius of bar/spiral, $R_\mr{co}$. Then the last term of the right hand side of equation (\ref{eq:rf}) is negative at $R$ around $R_\mr{co}$, thereby to compensate this export of gas due to the radial flow, $\sgmin$ becomes larger than that for the case without radial flow. In contrast, the advection term is positive in both the central and outer disk regions, so $\sgmin$ becomes lower. Then $\kin$ in the case of non-zero $v_R$ at $R < R_\mr{co}$ ($R > R_\mr{co}$) becomes smaller (larger) than that in the case of $v_R = 0$. Thus radial flow will provide non-monotonous changes in the relation between $\sgmin$ and $\sgmsf$. However when considering the effects of radial flow over the entire disk, the difference of $\kin$ between the case of zero and non-zero $v_R$ may be small, implying that $\sgmin$ is on average proportional to $\sgmsf$ even including the effect of radial flow. Thus the effect of radial flow on the derived properties of gas inflow is expected to be insignificant. 

The effect on the outflow rate, $\sgmout$, by the presence of radial flow can be inferred as follows. For the disk with the negative RMG, because metal-poor (rich) gas present in the outer (inner) disk regions flow inward (outward), the RMG tends to be flattened by such radial flow. Then $\kout$ may become smaller in response to this flattened RMG since the outflow with smaller $\kout$ works so as to steepen the RMG more efficiently, as mentioned in Section 4.1. 

%%% Sec.4.4.2 %%%%%%%%%%%%%%%%%%%%%%%%%%%%%%%%%%%%%%%%%%%%%%%%
\subsubsection{Re-accretion of metal-enriched galactic wind}
Recent hydro-dynamical simulations for galaxy evolution suggest that some fraction of gas ejected from the galactic disk by feedback processes remain in the host halo and accrete back into the disk (Oppenheimer \& Dav\'e 2008; Dav\'e et al. 2011). Moreover outflowing gas may acquire additional angular momentum via feedback processes and preferentially re-accrete into outer disk regions than the radii where they originally resided (Bekki et al. 2009). Therefore re-accretion of metal-enriched galactic wind, if it occurs, may play a role in transporting heavy elements from the inner to outer disk regions, thereby strongly affecting the radial distribution of heavy elements inside the galactic disk (e.g., Tsujimoto et al. 2010; Gibson et al. 2013). 

To investigate this influence on our results, we consider the case that inflowing gas consists of not only primordial one, but also re-accreting enriched one, for which we adopt a non-zero $Z_\mr{in}$ to solve equation (\ref{eq:dm2}). For this experiment, we assume that the metallicity of inflowing gas at each radius is proportional to the total metallicity of the galaxy, $Z_\mr{in} \tr = a(R) Z_\mr{tot} (t)$, where the proportional coefficient, $a(R)$, is assumed to increase linearly along the radius so as to represent re-accretion of metals preferentially into the outer disk regions. As a fiducial example, we adopt $a(R) = 0.1 + 0.1 (R/10 \;\mr{kpc})$ so that it changes from 0.1 at $R$ = 0 kpc to 0.2 at $R$ = 10 kpc. We believe that this case provides sufficiently large $Z_\mr{in}$ to examine the effect, because the actually observed metallicities in high velocity clouds presently falling into our Galactic disk are estimated as only 0.09 times the solar metallicity (e.g., Wakker et al. 1999).

We perform this calculation for both the sRMG and fRMG models and find that this form of non-zero $Z_\mr{in} \tr$ makes about 20 \% of metals ejected from the entire disk eventually fall back at later epochs. We also find that the results from these calculations are basically the same as those shown in Section 3: the change in the parameters $\eta_\mr{in/out}$ and $k_\mr{in/out}$ by adopting this non-zero $Z_\mr{in}$ are typically only about 10 \% and generally confined to less than 30 \% at all epochs for both the sRMG and fRMG models, so the redshift evolutions of these parameters remain the same as those shown in Figure \ref{fig:3}. Therefore we conclude that the effect of re-accretion of metal-enriched galactic wind does not significantly modify our conclusions in this paper.

%%% Sec.4.5 %%%%%%%%%%%%%%%%%%%%%%%%%%%%%%%%%%%%%%%%%%%%%%%%
\subsection{The Total Inflow and Outflow Rates}
The total inflow and outflow rates at each redshift are obtained by integrating the inflow and outflow profiles along the radius shown in Figure \ref{fig:1}. The information on these total gas flow rates, which are to be compared with the observations of distant MWLGs, can place useful constraints on galactic chemical evolution models. We here present the integrated inflow and outflow masses obtained from our analysis, and discuss their roles in galaxy evolution. 

%%% Sec.4.5.1 %%%%%%%%%%%%%%%%%%%%%%%%%%%%%%%%%%%%%%%%%%%%%%%%
\subsubsection{Inflow}
\begin{figure}
\figurenum{6}
\begin{center}
\includegraphics[width=8.0cm,height=5.0cm]{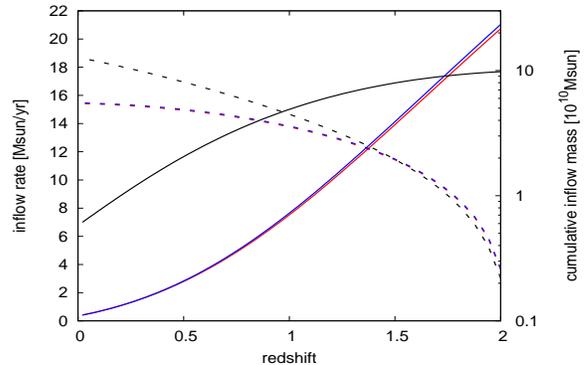}
\end{center}
\caption{The total inflow rate obtained by the integration of the inflow profile along the radius within $R$ = 10 kpc (solid line) and the cumulative inflow mass after $z = 2$ (dashed line) as a function of redshift. The red and blue lines show the results for the sRMG and fRMG models, respectively. For comparison, the case for the ideal baryonic accretion is shown with black lines.}
\label{fig:tin} 
\end{figure}

Figure \ref{fig:tin} shows the total inflow rates (solid lines) and cumulative inflow masses after $z = 2$ (dashed lines) derived by integrating the inflow profile within $R$ = 10 kpc, which is the outer boundary of the disk in our analysis as mentioned in Section 2, where the results from the sRMG and fRMG models are shown with red and blue lines, respectively. We find that the total inflow rates are highest at $z = 2$ and decrease rapidly with time at later epochs. It is clear that the total inflow rates for two models are remarkably similar, although the internal properties of gas inflow between the models are largely different. This result suggests that the total mass of inflowing gas into a galaxy is determined by the cumulative properties of the galaxy, such as its total stellar mass, SFR and mass budget of heavy elements, rather than the spatially resolved properties, such as the radial distributions of baryonic components and their metallicities. The total masses accreted on the galactic disk after $z = 2$ are $5.4 \times 10^{10} M_\odot$ and $5.5 \times 10^{10} M_\odot$ for the sRMG and fRMG models, respectively. These masses correspond to the present stellar disk mass in MWLGs. 

%%% Figure 6 %%%

In Figure \ref{fig:tin} we also show the ideal baryonic accretion into the halo of MWLGs with black lines, which is obtained by multiplying the cosmic baryonic fraction, $f_\mr{b} \sim 0.17$, by the average virial mass growth rate for MWLGs (Bouch\'e et al. 2010, and references therein). We find that this ideal accretion is clearly different from the actual inflows presented by our analysis, as the latter decline much faster than the former. This difference implies that not all the gas existing in the halo of MWLGs has been supplied on the galactic disk, in particular at lower redshifts. This suppression of gas infall may be caused by feedback from supernovae or active galactic nucleus in the galaxy, which makes halo gas heated up before accreting on the galactic disk. Recently, Lu et al. (2015) investigated the semi-analytic model, which includes the effects of preheating mechanisms, and found that the model can reproduce remarkably well various observational scaling relations such as the cold gas fraction - virial mass relation, disk size - stellar mass relation and its evolution. Thus the feedback processes in a galaxy may be linking up not only to the production of outflow, but also to the several properties of star-forming galaxies including baryonic accretion.

%%% Sec.4.5.2 %%%%%%%%%%%%%%%%%%%%%%%%%%%%%%%%%%%%%%%%%%%%%%%%
\subsubsection{Outflow}
\begin{figure}
\figurenum{7}
\begin{center}
\includegraphics[width=8.0cm,height=5.0cm]{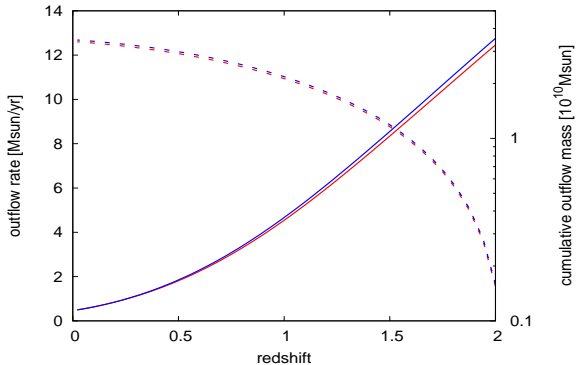}
\end{center}
\caption{The total outflow rate obtained by the integration of the outflow profile along the radius within $R$ = 10 kpc (solid line) and the cumulative outflow mass after $z = 2$ (dashed line) as a function of redshift. The red and blue lines show the results for the sRMG and fRMG models, respectively.}
\label{fig:tout1} 
\end{figure}

Figure \ref{fig:tout1} is the same as Figure \ref{fig:tin}, but for the total outflow rates. We find that these rates show nearly the same evolution for both models, i.e., declining rapidly with time, similarly to the total inflow rates. The total masses ejected from the galactic disk after $z = 2$ are $3.4 \times 10^{10} M_\odot$ and $3.5 \times 10^{10} M_\odot$ for the sRMG and fRMG models, respectively. 

\begin{figure}
\figurenum{8}
\begin{center}
\includegraphics[width=8.0cm,height=5.0cm]{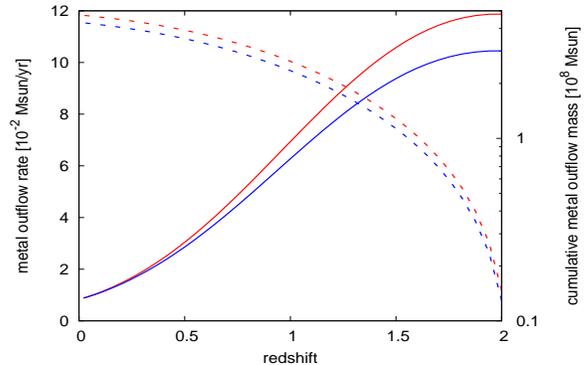}
\end{center}
\caption{The total metal outflow rate obtained by the integration of the metal outflow profile along the radius within $R$ = 10 kpc (solid line) and the cumulative metal outflow mass after $z = 2$ (dashed line) as a function of redshift. The red and blue lines show the results for the sRMG and fRMG models, respectively.}
\label{fig:tout2} 
\end{figure}

Figure \ref{fig:tout2} shows the total outflow rate of metals and their cumulative mass. This flow is also decreasing with time, but this decreasing rate is somewhat slower than that for gas outflow because the chemical abundance of the galactic disk is a increasing function of time. The total masses of metals ejected from the galactic disk after $z = 2$ are $4.7 \times 10^{8} M_\odot$ and $4.3 \times 10^{8} M_\odot$ for the sRMG and fRMG models, respectively. These masses correspond to about 40 \% of the mass of all metals synthesized in the star formation history of MWLGs, $Y (1 - \ret) M_\ast$. Similar results were reported by the previous works, which compared the mass of metals still contained in the galactic disk with that produced by all the star formation processes (e.g., Zahid et al. 2012). According to Peeples et al. (2014), about 80 \% of metals are already lost from the galactic disk in MWLGs probably due to outflow. Thus these works including our current analysis imply that galactic outflows have provided significant effects on the chemical evolution of galaxies.

%%% Figure 7 %%%

%%% Figure 8 %%%

%%% Sec.5 %%%%%%%%%%%%%%%%%%%%%%%%%%%%%%%%%%%%%%%%%%%%%%%%
\section{SUMMARY AND CONCLUSIONS}
We have investigated the gas inflow and outflow rate densities in Milky Way-like galaxies as a function of the SFR density, in order to get new insights into the baryonic processes tightly relating to galaxy formation and evolution. For this purpose, we solve equations (\ref{eq:dm1}) and (\ref{eq:dm2}) for the evolutions of the surface mass densities of gas and metals at each radius in a galactic disk, based on the observational studies of distant star-forming galaxies, such as the redshift evolution of their stellar mass distribution, their fundamental metallicity relation and the evolution of their radial metallicity gradients. The solutions of these equations allow us to evaluate the efficiency of inflow and outflow rate densities, $\sgmin$ and $\sgmout$, and their dependence on the SFR density parameterized  with $\eta_\mr{in/out}$ and $k_\mr{in/out}$ as presented in equation (\ref{eq:fit}). 

We have found that $\kin$ is approximately unity at all redshifts, representing the proportionality of the inflow rate to the SFR, and $\etain$ is a decreasing function of time. Such properties of gas inflow can be understood based on the equation of gas mass evolution. Therefore gas inflow dominates the budget of gas and star formation activity in a galactic disk.

For gas outflow, while $\etaout$ decreases with decreasing redshift, similarly to $\etain$, $\kout$ deviates largely from unity at most redshifts, indicating that gas outflow rate is not necessarily proportional to SFR. Moreover, we have found that the relation between the outflow rate and SFR strongly depends on the evolution of the adopted radial metallicity gradient. Thus galactic outflow provides the significant influence on the distribution of heavy elements in disk galaxies.

The relation between the outflow rate and SFR is understood based on the momentum and energy-driven wind models. It is found that momentum-driven (energy-driven) wind mechanisms can well reproduce the redshift evolution of $\kout$ based on the steepening (flattening) radial metallicity gradients with time. However, the detailed spatial distribution of metallicity in Milky Way-like galaxies has not been constrained well from the observations of distant galaxies. Future observations armed with the state-of-the-art surface-spectroscopic instruments such as VLT/KMOS will be important to unveil the detailed spatial distribution of metallicity in distant galaxies at several redshifts. If the well measured radial metallicity gradients and their evolution are available from such observations, then the method that we have presented in this work will be useful to constrain the main driving mechanism for their galactic outflows.

%%%%%%%%%%%%%%%%%%%%%%%%%%%%%%%%%%%%%%%%%%%%%%%%%%%%%%%%%
\acknowledgments
We are grateful for the referee for her/his constructive comments that have helped us to improve our paper. This work is supported in part by a Grants-in-Aid for Scientific Research from the Japan Society for the Promotion of Science (JSPS; No. 27-2450 for DT) and by a Grant-in-Aid for Scientific Research (25287062) of the Ministry of Education, Culture, Sports, Science and Technology in Japan.

%%%%%%%%%%%%%%%%%%%%%%%%%%%%%%%%%%%%%%%%%%%%%%%%%%%%%%%%%

\clearpage
%%%%%%%%%%%%%%%%%%%%%%%%%%%%%%%%%%%%%%%%%%%%%%%%%%%%%%%%%

\end{document}